\documentclass[11pt,twoside]{article}

\usepackage{asp2006}
\usepackage{epsf}

\begin{document}
\title{Bars in Local Galaxies: Evidence for a higher Optical Bar
  Fraction in Disk-Dominated Galaxies}   %%% Fill in title
\author{Fabio D. Barazza\altaffilmark{1,2}, Shardha
  Jogee\altaffilmark{1}, and Irina Marinova\altaffilmark{1}}
   %%% Fill in author names
\altaffiltext{1}{Department of Astronomy, University of Texas at Austin, 1 University
Station C1400, Austin, TX 78712-0259, USA}
\altaffiltext{2}{Laboratoire d'Astrophysique, \'Ecole
Polytechnique F\'ed\'erale de Lausanne (EPFL), Observatoire de Sauverny, CH-1290 Versoix,
Switzerland}    %%% Fill in author affiliations

\begin{abstract} %%% Abstract to run on from here.
We present a study of large-scale bars in the local Universe, based on a large
sample of $\sim3692$ galaxies, with $-18.5 \leq M_g < -22.0$ mag and redshift
$0.01 \leq z < 0.03$, drawn from the SDSS.
Our sample includes many galaxies that are 
disk-dominated and of late Hubble types. Both color cuts and S\'ersic
cuts yield a similar sample of $\sim2000$ disk galaxies. We
characterize bars and disks by ellipse-fitting  r-band images
and applying quantitative criteria. After excluding highly inclined
($>60^{\circ}$) systems, we find the following results. (1)~The optical
$r$-band fraction ($f_{opt-r}$) of barred galaxies is
$\sim48\%-52\%$. (2)~When galaxies are
separated according to normalized half light radius ($r_e/R_{24}$), a
remarkable result is seen: $f_{opt-r}$ rises sharply,
from $\sim40\%$ in galaxies that have small $r_e/R_{24}$ and
visually appear to host prominent bulges, 
to $\sim70\%$ for galaxies that have large $r_e/R_24$ and
appear disk-dominated. (3)~$f_{opt-r}$
rises for galaxies with bluer colors and higher central
surface brightness. A weaker rise is seen toward lower masses. (4) We
find that $\sim20\%$ of our sample
of disk galaxies appear to be ``quasi-bulgeless''. (5)
If we restrict our sample to bright galaxies and only consider bars
that are strong (ellipticity $\ge0.4$) and large enough (semi-major axis
$\ge1.5$ kpc) to be reliably characterized via ellipse-fitting out to
$z\sim0.8$, we get an optical $r$-band fraction for strong bars
$f_{\rm opt-s}$ of $\sim34\%$. This value is  higher only by a
modest factor of 1.4, compared to the value of $\sim24\%\pm4\%$
reported at $z\sim0.7-1.0$. If one assumes that the increasing obscuration
by dust and star formation over $z\sim0$ to 1.0 causes a further
artificial loss of bars, the data even allow for a constant or rising fraction
of strong bars with redshift.
\end{abstract}

%%% MAIN BODY OF TEXT GOES HERE. CONSULT "INSTRUCTIONS FOR AUTHORS USING
%%% LATEX2E MARKUP", SECTIONS 2.3-2.6 FOR HELP WITH EQUATIONS, FIGURES,
%%% AND TABLES.

%\section{}   %%% Top level section head (remove "%" symbol)
%\subsection{}   %%% Second level section head (remove "%" symbol)
%\subsubsection{}   %%% Lowest level section head (remove "%" symbol)
\section*{Introduction}    %%% Unnumbered top level section head (remove "%" symbol)
Bars are believed to be very important with regard to the dynamical and secular
evolution of disk galaxies, particularly in redistributing the angular momentum
of the baryonic and dark matter components of disk galaxies. From a
theoretical perspective, it is possible to model some aspects of the
evolution of disks and bars, and their interactions.
However, it remains unclear why a specific galaxy has a bar, but a seemingly
similar galaxy is unbarred; or why some barred galaxies have a classical bulge,
whereas others harbor a disky bulge. In our study, we use
a sample of $\sim2000$ disk galaxies, at $z=0.01-0.03$ with $M_r \sim
-18.5$ to $-22.0$ mag. The large sample allows for the first time to
have $100-200$ galaxies per bin, while binning galaxies in terms of
different host galaxy parameters (e.g. luminosity, color, surface brightness).
\subsection*{Results}   %%% Unnumbered second level section head (remove "%" symbol)
After excluding highly inclined ($>60^{\circ}$) systems, we find:
\begin{enumerate}
\item 
The average optical $r$-band bar fraction ($f_{\rm opt-r}$) in our sample,
which primarily consists of late-type disk-dominated galaxies, is $\sim48\%-52\%$.

\item 
The optical $r$-band fraction rises sharply, from $\sim40\%$ in galaxies
that have small $r_{\rm e}$/$R_{\rm 24}$ (or large
bulge-to-disk ratio) and visually appear bulge-dominated, to
$\sim70\%$ for galaxies that have large $r_{\rm e}$/$R_{\rm 24}$
(Fig. 1a).

\item 
Similar trends in the optical bar fraction  
are found using the central surface brightness (Fig. 1b) and color.
Bluer galaxies have higher  bar fractions ($\sim58\%$ at $g-r=0.3$) than
the redder objects ($\sim32\%$ at $g-r=0.65$).

\item 
We find that $\sim20\%$ of the moderately inclined disk galaxies appear to
be ``quasi-bulgeless'',  without a classical bulge. 

\item
If we restrict our sample to bright galaxies and only consider bars
that are strong (ellipticity $\ge0.4$) and large enough (semi-major axis
$\ge1.5$ kpc) to be reliably characterized via ellipse-fitting out to
$z\sim0.8$, we get an optical $r$-band fraction for strong bars
$f_{\rm opt-s}$ of $\sim34\%$. This value is  higher only by a
modest factor of 1.4, compared to the value of $\sim24\%\pm~4\%$
reported at $z\sim0.7-1.0$.
Assuming further artificial loss of bars caused by an increased obscuration
by dust and star formation over $z\sim$~0 to 1.0, the data even allow
for a constant or rising fraction of strong bars with redshift.
\end{enumerate}
\begin{figure}
\plottwo{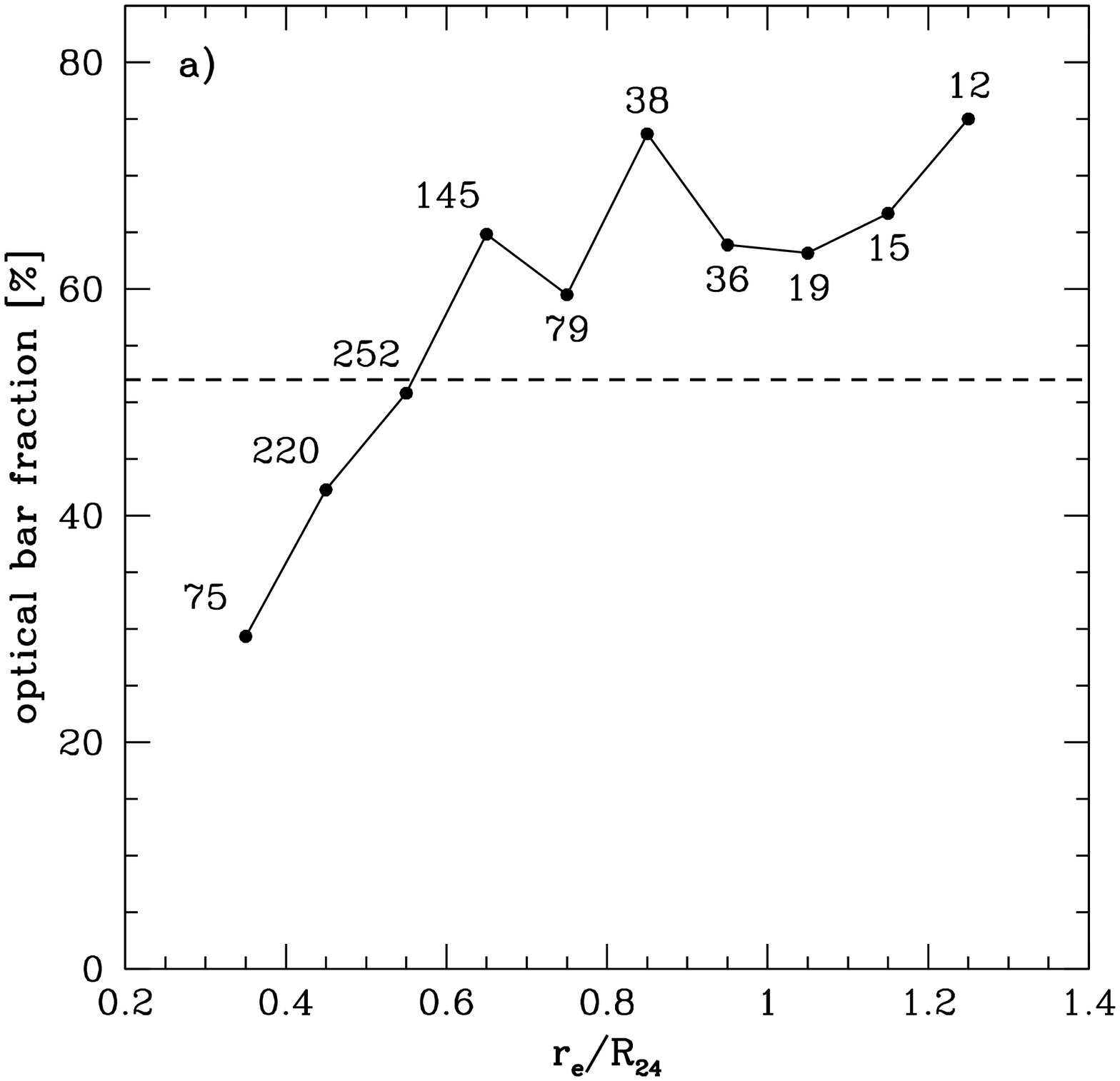}{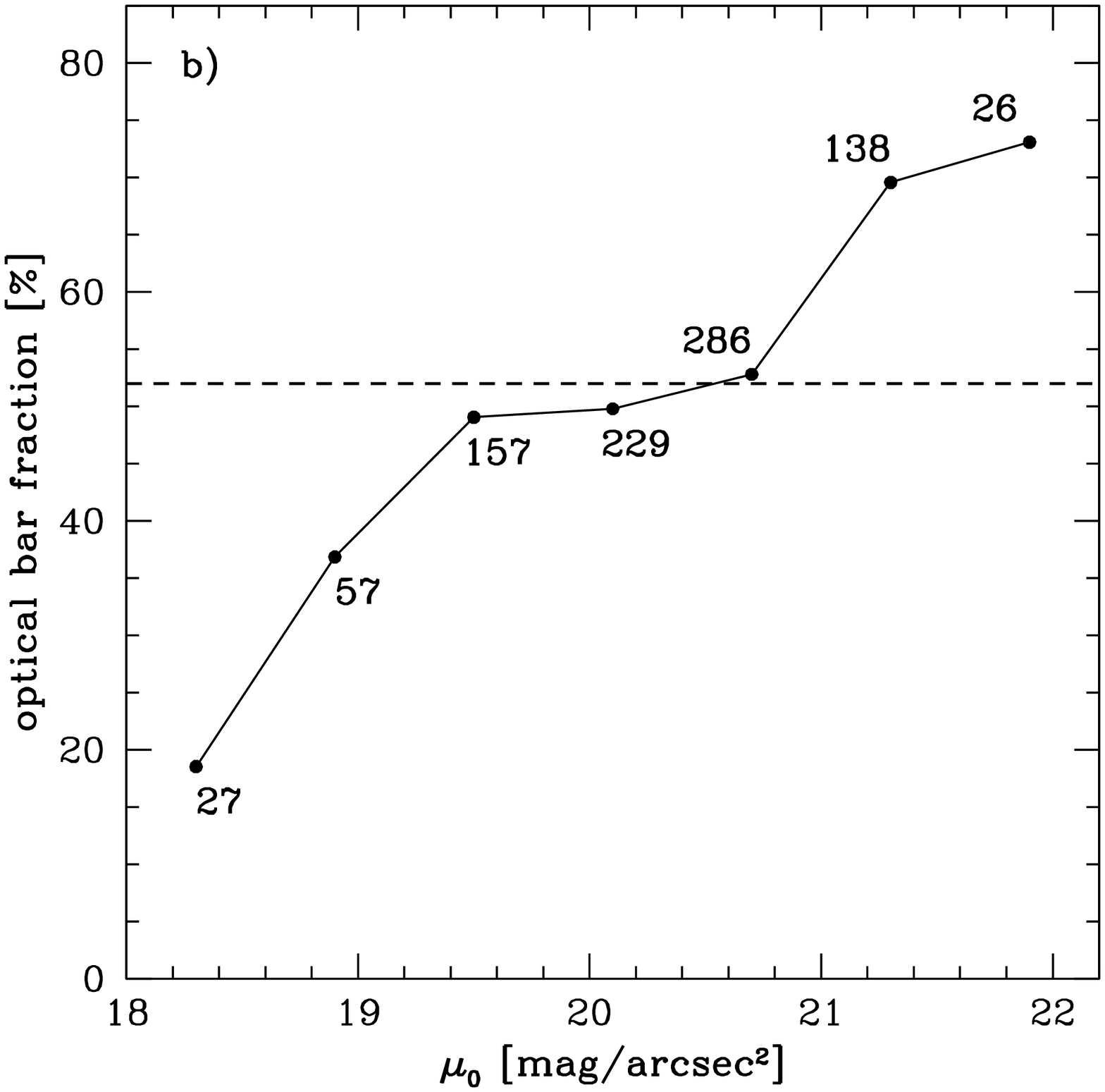}
\caption{{\bf a)} The optical bar fraction as a function of
$r_{\rm e}$/$R_{\rm 24}$.  The most compact galaxies have a bar
fraction of $\sim30\%$, whereas the most extended galaxies reach a
value of $\sim70\%$. {\bf b)} The optical bar fraction as a function of central
surface brightness. In both panels the numbers next to the points
indicate the number of galaxies in the corresponding bins. The dashed lines
indicate the total optical bar fraction ($52\%$).}
\end{figure}

\acknowledgements %%% Text of acknowledgements runs on after this command.
F.D.B, S.J, and I. M. acknowledge support from the National Aeronautics
and Space Administration (NASA) LTSA grant NAG5-13063, NSF grant AST-0607748,
and HST-GO-10861.

%%% THE BIBLIOGRAPHY
%%%
%%% CONSULT SECTION 3 OF "INSTRUCTIONS FOR AUTHORS" FOR HOW TO USE NATBIB.
%%% AUTHORS ARE ENCOURAGED TO USE EITHER THE "THEBIBLIOGRAPY" ENVIRONMENT
%%% BY UNCOMMENTING (DELETING THE "%" SYMBOL) THE COMMANDS BELOW, OR BY
%%% USING THE BIBTEX ENVIRONMENT. TO FIND OUT WHICH IS APPLICABLE TO YOUR
%%% CONTRIBUTION, CONSULT THE VOLUME EDITORS FOR YOUR PROCEEDINGS.
%%%

%\begin{thebibliography}{}
%\bibitem[]{}
%\bibitem[]{}
%\bibitem[]{}
%\bibitem[]{}
%\bibitem[]{}
%\bibitem[]{}
%\bibitem[]{}
%\bibitem[]{}
%\bibitem[]{}
%\bibitem[]{}
%\bibitem[]{}
%\bibitem[]{}
%\end{thebibliography}

\end{document}